%% file: main.tex
\documentclass[journal]{IEEEtran} 

\usepackage{graphicx}
\usepackage{epsfig}
\usepackage{tabularx}
\usepackage{amssymb}
\usepackage{latexsym}
\usepackage{amsmath}
\usepackage{algorithm}
\usepackage[noend]{algpseudocode}
\usepackage{color}
\usepackage{relsize}
\usepackage{mathtools}
\usepackage{amsthm}
\usepackage{cite}
\usepackage{booktabs}
\usepackage{cite}

\newcommand{\RN}[1]{%
  \textup{\uppercase\expandafter{\romannumeral#1}}
  }

\IEEEoverridecommandlockouts
\IEEEpubid{\makebox[\columnwidth]{978-1-7281-9023-5/21/\$31.00~ \copyright{} 2021 IEEE \hfill}
\hspace{\columnsep}\makebox[\columnwidth]{ }}

\begin{document}

%\title{Analyzing 5G Radio Access Networks for Smart Grid Performance via a Co-Simulation Environment}
\title{Power Systems Performance under 5G Radio Access Network in a Co-Simulation Environment}

% \author{Rahul Iyer$^*$\thanks{*Both the authors contributed equally to this work.\\ (Email:\{irahulrajan, biplavc, mehrizi, vijays}@vt.edu),
%         Biplav Choudhury$^*$,
%         Ali Mehrizi-Sani and
%         Vijay K. Shah

% }

\author{Rahul Iyer$^{1*}$\thanks{* Both first authors contributed equally to this work. \\ 
This work is supported in part by the National Science Foundation under awards ECCS-1953198 and ECCS-1953213 and in part by the Commonwealth Cyber Initiative, an investment in the advancement of cyber R\&D, innovation, and workforce development (www.cyberinitiative.org).

This paper has been accepted for publication in IEEE ISIE 2021. This is a preprint version of the accepted paper.
},
        Biplav Choudhury$^{2*}$, Vijay K. Shah $^2$ 
        and Ali Mehrizi-Sani$^1$ 
        \\
        $^1$Power and Energy Center, $^2$Wireless@Virginia Tech \\ Bradley Department of Electrical and Computer Engineering, Virginia Tech, Blacksburg, VA. USA \\
        Emails:\{irahulrajan, biplavc, vijays, mehrizi\}@vt.edu
}

\maketitle

\begin{abstract}

% In this project, we consider smart grid applications that use a 5G network as means of communication among the components for system control. We investigate the affect of scheduling in terms of Resource Blocks (RB) in the performance of the smart grid. The performance is compared with ideal simulation based results and the inference includes comments on whether the Smart Grid cases considered would meet the stability requirements, when a 5G communication network is involved.

Communication can improve control of important system parameters by allowing different grid components to communicate their states with each other. % about their %respective states. 
This information exchange requires a reliable and fast communication infrastructure. 5G communication can be a viable means to achieve this objective. This paper investigates the performance of several smart grid applications %that involve distributed control 
under a 5G radio access network. %This study offers key insights on 5G's capability to support critical functions. 
Different scenarios including set point changes and transients are evaluated, and the results indicate that the system maintains stability when a 5G network is used to communicate system states.

\end{abstract}


\IEEEpeerreviewmaketitle

\input{1-introduction}

\input{2-cosimulator}

\input{3-system-model}

\input{4-pp}

\input{5-cspm}

\input{6-conclusion}
% \input{analysis} % give exact setting of the system model here.

% \section*{Acknowledgment}

\ifCLASSOPTIONcaptionsoff
  \newpage
\fi

%\newpage
% \bibliographystyle{IEEEtran}
% \bibliography{mybib}
% \bibliographystyle{unsrt}
\bibliographystyle{IEEEtran}
% \bibliography{mybib}
\bibliography{bibtex/bib/IEEEabrv.bib,mybib}

\end{document}

%% file: 1-introduction.tex
\section{Introduction}
\label{section - Introduction}

5G is the most recent wireless communication standard that improves rates and reliability over older generations such as 4G/LTE. Prominent applications of 5G communications include remote mining, vehicular communications, healthcare, and military \cite{Fivetop51:online}. All these applications rely on the extremely high data rates and reliability supported by 5G. In general, 5G applications are categorized into three main types based on their performance requirements: (i) \textit{enhanced mobile broadband (eMBB)}: eMBB involves providing 5G connectivity to traditional cellphones which is expected to be its most commonly used application, (ii) \textit{ultra reliable low latency communications (URLLC)}: URLLC is designed for critical applications where reliability is extremely important, e.g., driving or robotic surgery, and (iii) \textit{massive machine type communications (mMTC)}: mMTC is designed for massive internet of things (IoT) devices where battery performance is more important than data rates. % which is expected to be the driving force behind industrial automation. 

%The different characteristics of these applications are described in Fig. \ref{fig:application} \cite{The5GPer50:online}.

% \begin{figure} [htb]
%     \vspace{-0.15 in}
%     \centering
%     \includegraphics[scale=0.65, trim={2cm 0cm 0cm 0cm},clip, angle=0]{images/types.pdf}
%     \vspace{0in}
%     \caption{5G application types \cite{The5GPer50:online}}
%     \label{fig:application}
%     \vspace{-0.05 in}
% \end{figure}

5G also has applications in control of the modern power system, often referred to as the smart grid. Smart grid leverages communication techniques to collect and distribute information about the system measurements to various grid components and the control center. For example, real-time feedback on electricity consumption can be used to dynamically adjust energy generation with the objective of improving energy efficiency \cite{foster2012results}. Traditionally, communication in power grid is designed using optical fibers which can be expensive as the grid becomes larger due to the large-scale connectivity needed between all the components involved \cite{Microsof15:online}. Therefore, there is a timely opportunity for a use case for providing connectivity to power grid components %where both the aspects of power transmission and distribution can be supported 
through 5G. %Network slicing in 5G is a particularly relevant application with respect to power grids due to its ability to assign independent slices in charge of different aspects of grid control \cite{5gnetwor51:online}. 5G network slicing is an ideal choice to enable smart grid services. Network slicing divides the 5G network into logically separated networks, where each one can be seen as a slice and allows 5G network slicing allows the power grid to flexibly customize specific slices with different network functions and different SLA (Service Level Agreement) assurances according to the needs, to meet different network requirements of various services mentioned above
Third-generation partnership project (3GPP) %, which is the standards organizations in charge of developing protocols for mobile telecommunications, 
recently formed a study group to standardize the end-to-end architecture of 5G networks specifically built for aiding smart grid operations and is expected to be formalized by Release 18\cite{3GPPR18L14:online}.

There has been a recent focus on investigating the benefits that connectivity can bring in smart grids. The authors in \cite{garau20175g} compare the response of a fault management system %, e.g., circuit breaker that intervenes after a fault has been identified with
4G-LTE and 5G systems. The results indicate that machine-to-machine connectivity abilities of 5G provide significant performance improvement compared with 4G. Reference \cite{dragivcevic2019future} reviews several applications that 5G can enable in smart grids and analyzes the associated challenges. In \cite{8715864}, a network servicing both eMBB and URLLC applications is considered, and its ability in handling line protection through URLLC is analyzed. The results show that provided the network coverage is above a certain threshold, differential line protection can be supported by the network. However, these research studies consider only a black box view of the 5G network without considering the implications of its various subsystems on the overall performance. This paper tries to address this gap. %Specifically, we consider the 5G 5G radio access network (RAN) 

This paper implements a co-simulation environment for power system and communication networks and employs this in an example application for distributed control. Distributed control techniques in smart grids involve distributed energy resources (DER) that coordinate for higher reliability and performance~\cite{6870484}, especially where centralized control is impractical. This can happen due to a large number of DERs, privacy and security requirements, and computational implications. Distributed control can be applied at different levels, including primary~\cite{1705638} and secondary~\cite{cspmref}.

The contributions of this work are

% While the Radio Access Network (RAN) and the Core Network (CN) together comprise an ideal and deployed 5G network, the connection between the RAN and CN is often implemented using fiber-optic cables due to which it doesn't show much variability. This implies that the RAN is the critical component in terms of 

\begin{itemize}
\item A Python-based 5G radio access network (RAN) simulator is developed to interface with general purpose power systems simulators such as PSCAD and MATLAB, thereby creating a co-simulation environment where the power grid components can communicate with each other through 5G links. This allows us to simulate the effects of RAN resource allocation on the smart grid. The RAN simulator is robust in terms of both interfacing with the different software environments and the number of devices that communicate within the power system.

\item The suitability of 5G in supporting the specific use cases of a power park and coordinated set point modulation is investigated using this co-simulation environment.

\item The performance of the smart grid under a 5G network is compared with an ideal scenario where the power grid components can communicate instantaneously.

\end{itemize}
The rest of the paper is organized as follows: Section~\ref{sec:cs} describes the overall design of the co-simulator and the interactions between the power and communication systems. Section~\ref{section:comm_system} describes the 5G communication simulator in more detail. Sections~\ref{sec:pp} and~\ref{sec:cspm} present two power system use cases and show their performance in the presence of a 5G RAN. Finally, Section~\ref{sec:conclusion} concludes the paper.

%% file: 2-cosimulator.tex
% \Vijay{A better approach may be to have a section named "Co-simulator design". And you explain how 5G communication system and power grid system interact with each other. And, then, you can go into the details of implementation of 5G communication system and power system (if needed).}

% \Vijay{Biplav - Include a nice end-to-end 5G network picture, with UEs as DERs, gNode and Core. And, distinguish that we focus on 5G RAN (i.e., gNodeB) in this work. }

% \Vijay{Rahul - I think it may be a good idea to have a general ``microgrid'' system to explain the 5G communications for Power Grid, before you go into specific use cases of Power Park and SPM. What do you think?

%\Rahul{Hi professor, I see your point but what exactly would that system include. We cannot add any tests on that system so what would be the content of that section ?}

%This section will explain the interaction between the power system and the 5G communication network. 
\section{Co-Simulator Design}
\label{sec:cs}
The objective of developing the co-simulation environment is to investigate the performance of a smart grid in the presence of a 5G RAN. The power system scenarios are simulated in %power system 
software tools PSCAD and MATLAB, and the communication environment is implemented through a Python script.

% \textcolor{red}{PSCAD is a graphical user interface that can be used for carrying out power systems electromagnetic transient simulations. Electrical power system modeling and simulations are also carried out using the component libraries under Simscape Electrical module in MATLAB. It can be used to design control systems for different power system scenarios.}

% Ideally the devices, i.e., distributed energy resources (DER) simulated in a smart grid communicate instantaneously. To evaluate the performance with 5G, 

\begin{figure} [htb]
    % \vspace{-0.15 in}
    \centering
    \includegraphics[scale=0.33, trim={0cm 6.2cm 5cm 0cm},clip, angle=0]{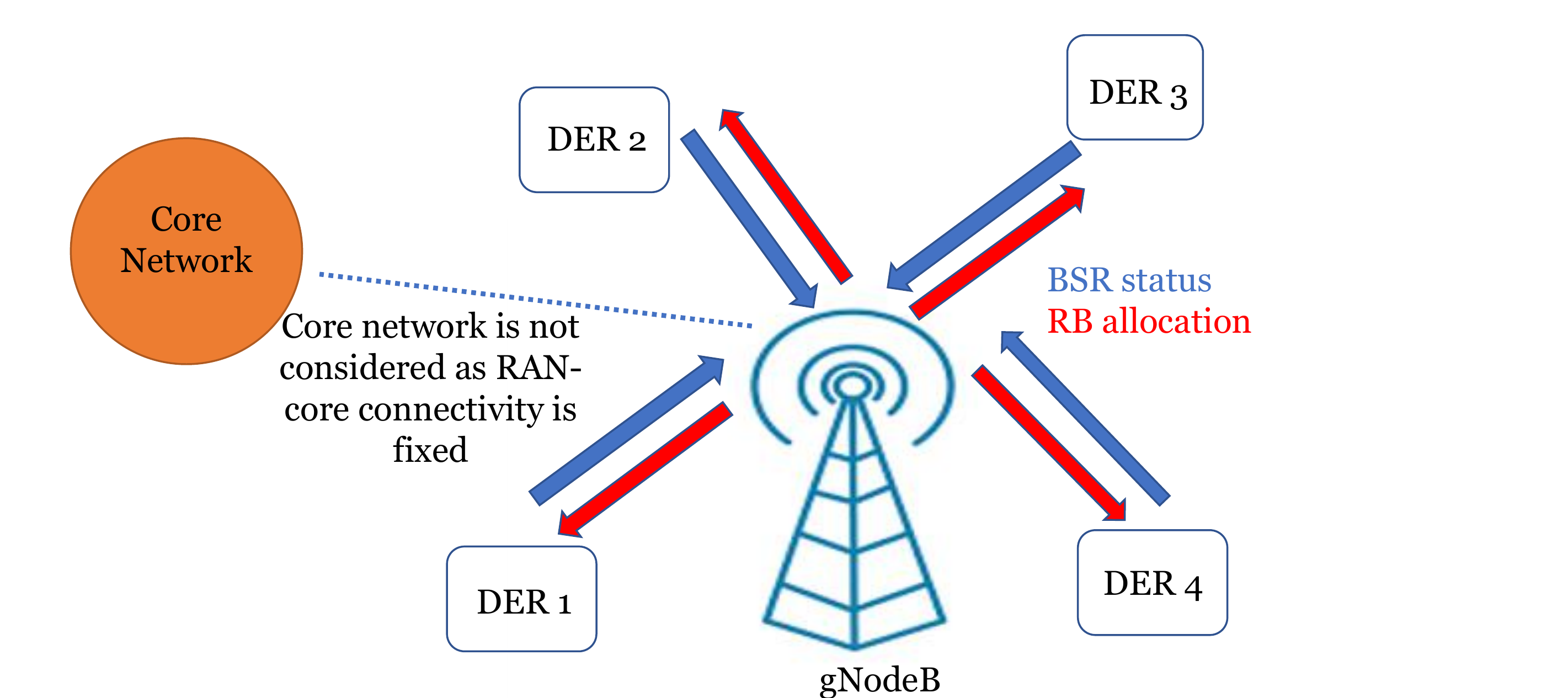}
    \vspace{0in}
    \caption{Schematic diagram of a 5G Radio Access Network}
    \label{fig:resource}
    % \vspace{-0.22 in}
\end{figure}

The Python script models the 5G communication between the smart grid DERs. Consider the scenario shown in Fig. \ref{fig:resource}. The DERs generate packets containing their state information, i.e., their local measurements, which need to be transmitted to the other DERs depending on the application. This transmission can be done when the DERs are allocated resources by the 5G Base Station (gNodeB), where these resources are the means to support the communication links. In 5G, these resources are termed resource blocks (RB) and RBs are allocated based on the number of packets a DER wants to transmit, which is known as that DER's buffer status report (BSR).  %One RB contains 12 sub-carriers in frequency domain with flexible spacing between the sub-cariers. %The time is synchronized between the power and communication systems periodically. 
Co-simulation operates as described below:

\begin{itemize}

\item Each DER %simulatedF in MATLAB and PSCAD 
transmits its BSR status to the Python-based gNodeB at an interval of $\tau$. This allows gNodeB to get the buffer status, i.e., packets pending to be sent from each DER.

\item Based on BSR, gNodeB allocates RBs to DERs using a pre-defined resource allocation policy $\pi$ at each transmission time interval (TTI). Based on the allocated RBs, DERs %in MATLAB and PSCAD 
are allowed to transmit their information, i.e., pass their values to the Python simulator. This information exchange between PSCAD and Python script is done through writing the values to a text file. % \cite{nasiriani2013embedded}. 

\item At each TTI, time is synchronized between the communication and power systems by allowing both of them to simultaneously step through the same duration of time.

\end{itemize}

\begin{figure} [htb]
    % \vspace{0 in}
    \centering
    \includegraphics[scale=0.35, trim={5.5cm 2.8cm 0cm 0cm},clip, angle=0]{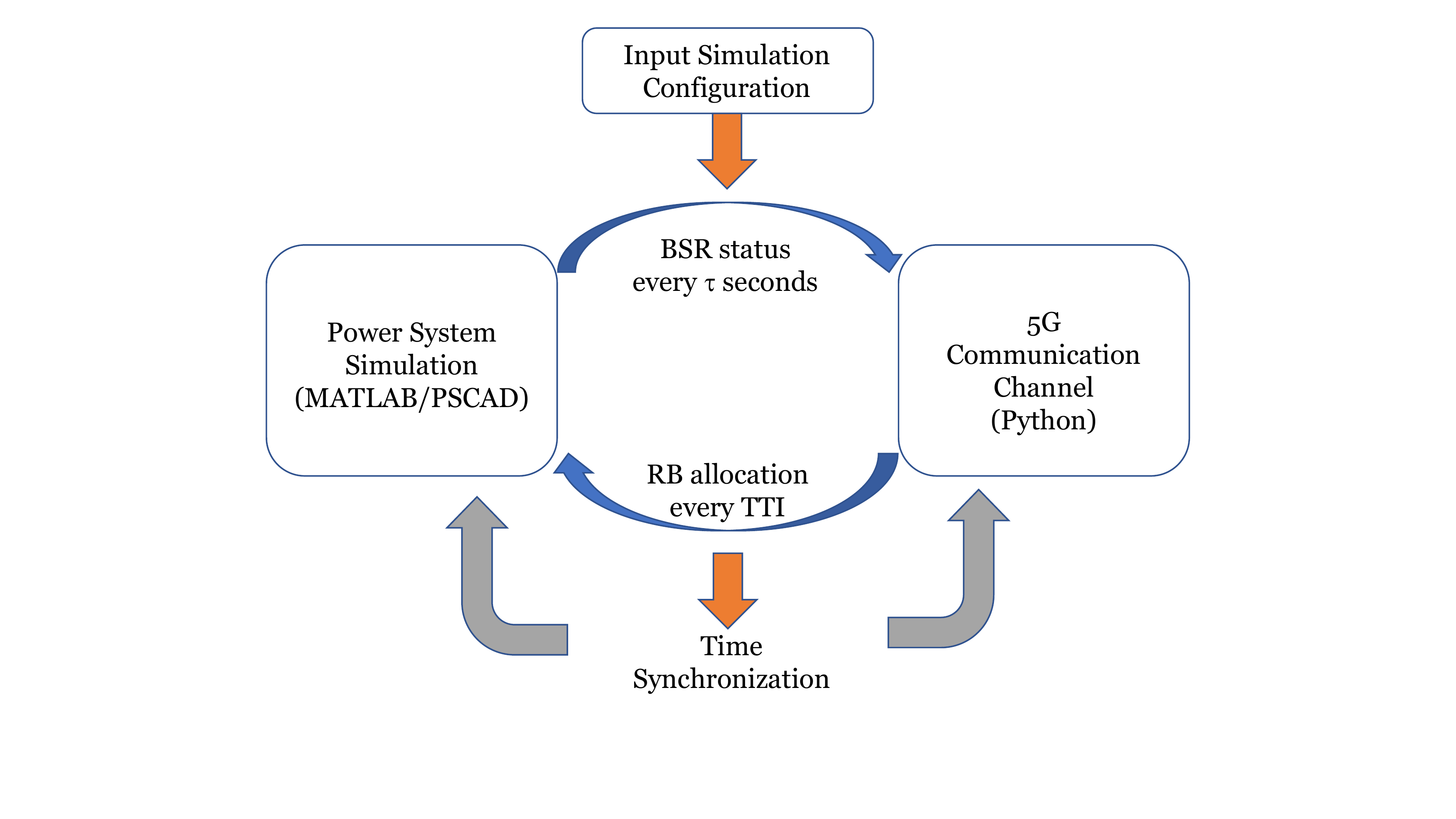}
    \vspace{0in}
    \caption{Design of the proposed co-simulator.}
    \label{fig:cosimulator}
    % \vspace{-0.2 in}
\end{figure}

% More details about the system will be presented in the upcoming sections. 
The general design of the co-simulator is shown in Fig.~\ref{fig:cosimulator}. More details are discussed in Section~\ref{section:comm_system}. %In the next section, we describe the 5G simulator in more detail. 

%% file: 3-system-model.tex
% \section{System Model}

\section{Communication System Design}
\label{section:comm_system}

A round-robin policy $\pi$ is used in our case where equal number of RBs are granted to each DER in a circular fashion, i.e., RBs are allocated to each DER in turn. Additionally, each DER transmits channel state information reference signal (CSI-RS) to gNodeB which is an indicator of channel quality. This indicator is called channel quality index (CQI) and provides information about the modulation order that can be used, where modulation order is the number of symbols that can be transmitted. Better CQI allows the use of higher modulation order. The channel is modeled as a random time-varying channel which results in varying modulation orders over the allocated RBs, where higher modulation order means higher throughput. %Based on all the factors discussed above, 
%The throughput attained by a DER is calculated using \eqref{eqn:throughput}. 
$M$ RBs are considered and based on the RB allocation, the throughput achieved in Mbps in each of the communication link is calculated as per 3GPP TS 38.306 \cite{43:online}. This is shown in \eqref{eqn:throughput}.   %using \eqref{eqn:throughput}.

\begin{equation}
% \nonumber
    10^{-6} \sum_{j=1}^{J} (v_{\text{Layers}}^{(j)}Q_{m}^{(j)}f^{(j)}R_{\text{max}}\frac{12N_{PRB}^{BW(j),\mu}}{T_{s}^{\mu}}(1-OH^{j}))
    \label{eqn:throughput}
\end{equation}

\noindent Here $J$ is the number of carriers aggregated in a carrier aggregation scenario. In 5G, up to 16 carriers can be aggregated. % which provides higher bandwidth and results in higher throughput. 
$v_{\text{Layers}}^{j}$ is the maximum number of layers in the $j$th component carriers (CC). It is also the number of streams and is restricted by the number of antennas used. $Q_{m}^{(j)}$ defines the modulation order used which depends on CQI. $f^{(j)}$ is the scaling factor used to scale throughput for various CC combinations. $R_{\text{max}}$ is the maximum coding rate and is typically set to $\frac{948}{1024}$ \cite{TS13821426:online}. $N_{\text{PRB}}^{BW(j),\mu}$ is the number of RBs allocated to a single DER. $\mu$ defines the 5G numerology selected, and this numerology decides the symbol time, $T_{s}^{\mu}$, calculated as $T_{s}^{\mu}=\frac{10^{-3}}{14.2^{\mu}}$. %The numerology $\mu$ is a key parameter and it decides important aspects like subcarrier spacing. 
Finally, the overhead $OH(j)$ for carrier $j$ is decided by the frequency band used (FR1 or FR2). %$R_{max}$ depends the coding type \cite{143:online}, in our case Low Density Parity Check (LDPC). 
gNodeB and DERs communicate in the the FR1 band \cite{9074973} (410 MHz to 7.125 GHz) using a bandwidth of $B$. Bandwidth is the range of radio wave frequencies allocated from the FR1 band. This achieved throughput decides the instants of information receptions at the DERs. Simulation parameters are shown in Table \ref{table:throughput_params}.

The BSR and CQI reports are assumed to be sent out of band (without the need of resources for transmission) and are assured error-free reception at the gNodeB \cite{NRTDDSym7:online}. Therefore the gNodeB always knows the BSR and CQI at each TTI. The 5G RAN is connected to a core network (CN). %The latency and reliability of this RAN-CN connection are usually fixed and do not vary~\cite{8715864}. 
As CN is a fiber cable-based network, the delay and reliability of this RAN-CN connection is fixed and do not vary~\cite{8715864}.
%~\footnote{Note that Release 17 of 5G cellular network propose Integrated and Access Backhaul (IAB) capability that will support wireless links between RAN and Core Ne.  
Therefore, we do not model the CN and instead focus on the RB allocation in the RAN network. As a result, the network performance is investigated from the perspective of RB allocation.

\begin{table}[htb]
 \centering
 \vspace{0 in}
 \caption{Parameters used to set up the simulation}
 \label{table:throughput_params}
%  \begin{tabular}{|p{3cm}|p{3.6cm}|}
    \begin{tabular}{ll}
    \toprule
    \textbf{Parameter}  & \textbf{Value} \\ \midrule
    Aggregated carriers $J$ & 2 \\ 
    Modulation order $Q_{m}^{(j)}$ & 2, 4, 6, 8 \\ 
    Maximum layers for $j$th carrier $v_{\text{Layers}}^{(j)}$  & 2 \\ 
    Scaling factor $f^{(j)}$ & 0.8 \\ 
    Numerology $\mu$ & 2 \\ 
    Number of RBs $M$ & 3 \\ 
    % RBs in downlink $N$ & 3 \\ 
    RBs allocated per DER $N_{\text{PRB}}^{\text{BW(j)}}$ & 1 \\ 
    Scheduling policy $\pi$ & Round robin \\ 
    BSR periodicity $\tau$ & 1 ms \\ 
    Transmission time interval TTI & 1 ms \\ 
    % Uplink frequency $f_{U}$ & 2.51 GHz \\ 
    Carrier frequency $f_{D}$ & 2.63 GHz \\ 
    Bandwidth $B$ & 5 MHz \\ 
    Packet size $L$ & 150 Bytes \cite{Ericsson66:online} \\ 
    \bottomrule

 \end{tabular}
 \vspace{0 in}
\end{table}

%  \begin{table}[htb]
%  \centering
%  \vspace{0 in}
%  \caption{System Performance for Power Park Case}
%  \label{table:pp_perform}
%   \scriptsize
% %  \begin{tabular}{|p{3cm}|p{3.6cm}|}
%     \begin{tabular}{lll}
%     \toprule
%     \textbf{Case}  & \textbf{Overshoot (in V)} & \textbf{Settling Time (in s)} \\ \midrule
%     Ideal  & 0.01 & 0.0111 \\ 
%     Communication & 0.103 & 0.01419 \\ \bottomrule
   
% \end{tabular}
% \vspace{0 in}
% \end{table}

%% file: 4-pp.tex
% \section{Use Cases}
% \label{sec:pp}

% The two cases namely $1.$ Power park and $2.$ Coordinated set-point modulation, were subjected to different scenarios and their performance was analyzed in the presence of a 5G communication network.

% The following indicators are used to evaluate the performance of the test cases:
% \begin{itemize}
%     \item \textbf{Overshoot:} It is a measure of the maximum deviation of the system response in event of an external disturbance or a step change in the set point. If $x_{max}$ denotes the peak system response to a set point change $x_{sp}$, Overshoot is defined as 
%     \begin{equation*}
%         x_{os}= \mid\frac{x_{max}-x_{sp}}{x_{sp}}\mid
%     \end{equation*}
%     \item \textbf{Settling Time:} The time taken for the system response $x(t)$ that strays away from a error band $\mid\epsilon\mid$ to settle down and stay within the error band is called the settling time. For the cases considered under this project, this band is set at 3\% 
% \end{itemize}

\section{Use Case I: Power Park} 
\label{sec:pp}
\subsection{System Description}

\begin{figure} [htb]
    \centering
    \includegraphics[scale=0.4]{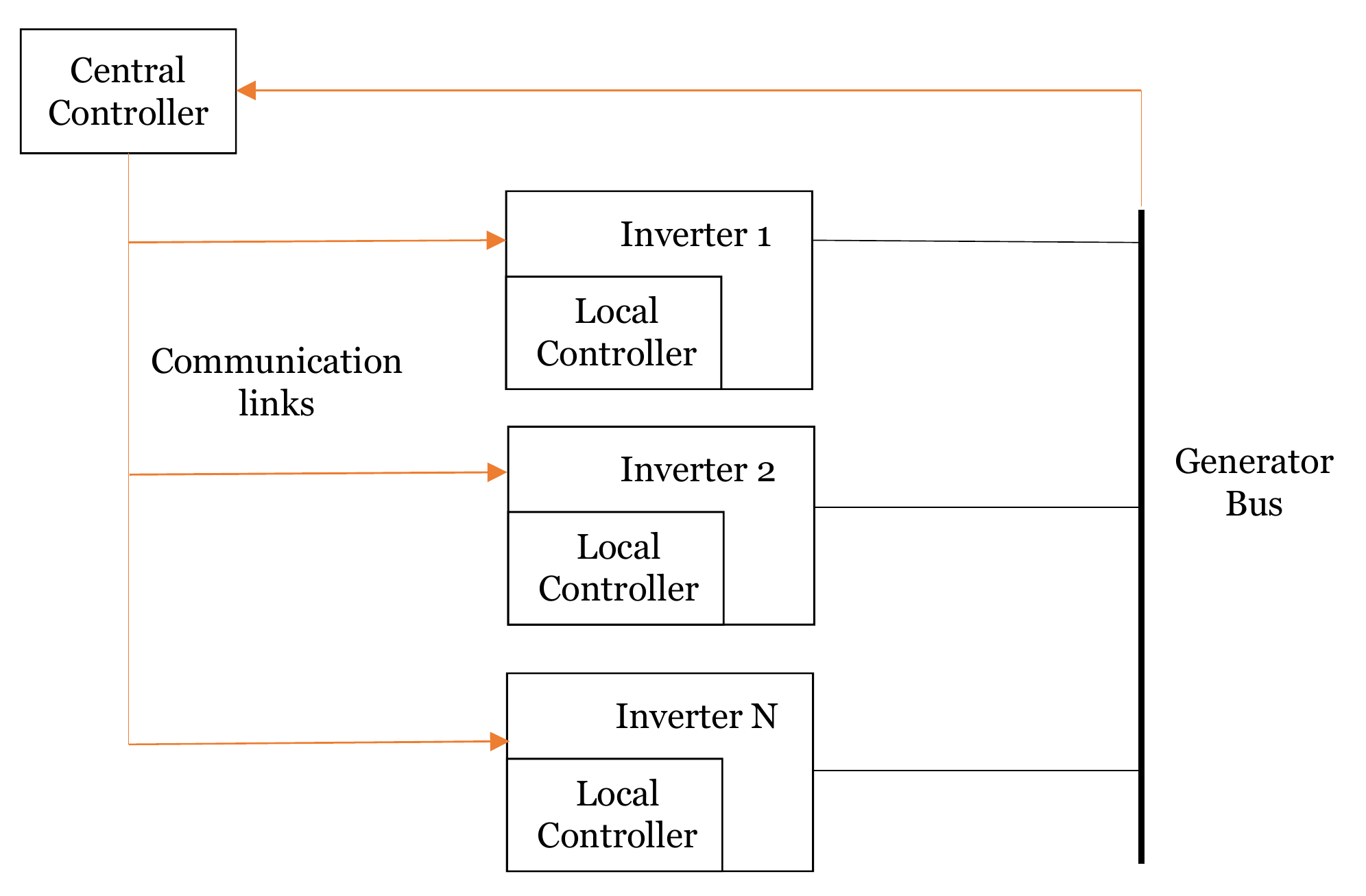}
    \caption{Power park system model.}
    \label{fig:fr_pt}
    %\vspace{-0.15in}
\end{figure}

This case involves distributed control of $N$ inverters connected to a common bus \cite{1705638}, as shown in Fig.~\ref{fig:fr_pt}. %Fig. \ref{fig:pp_block_diag} shows this concept.
%   \begin{figure} [htbp]
%      \vspace{0.15 in}
%      \centering
%      \includegraphics[scale=0.35, trim={0cm 0cm 0cm 0cm},clip, angle=0]{pp_b1.pdf}
%     \vspace{0in}
%      \caption{Modular arrangement for inverters \cite{1705638}.}
%      \label{fig:pp_block_diag}
%      \vspace{0.05 in}
%  \end{figure}

\begin{figure} [htbp]
    \centering
    \includegraphics[width=\columnwidth]{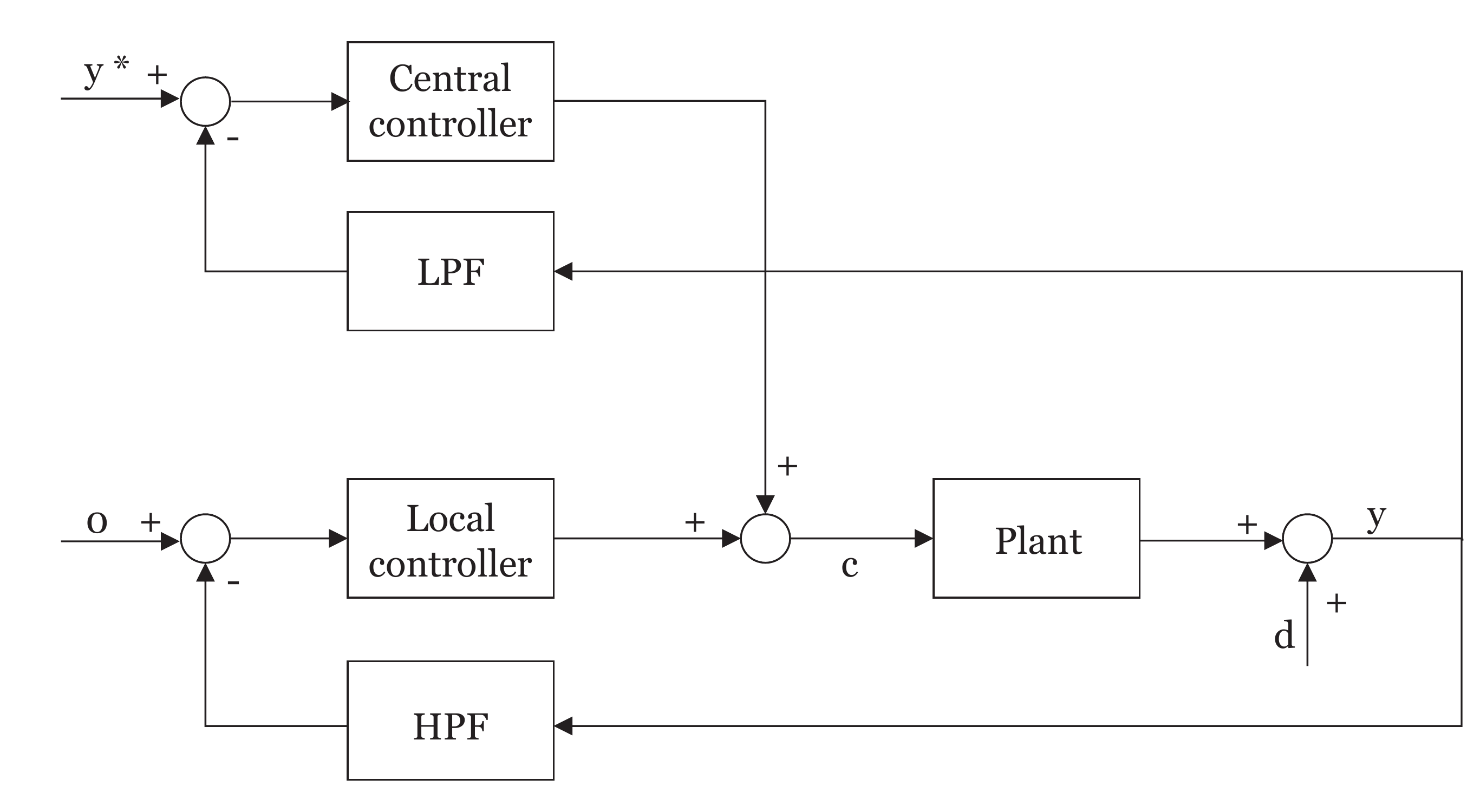}
    \caption{Frequency partitioning for inverter control.}
    \label{fig:freq_part}
    %\vspace{-0.15in}
\end{figure}

This case employs frequency partitioning, which divides the control tasks between the remote central controller and the distributed local controllers of each inverter. The concept of frequency partitioning is shown in Fig.~\ref{fig:freq_part}. This method reduces the required communication bandwidth between the remote central controller and the local controllers. The low frequency component of the control signal is provided by the central controller, while its high frequency component is generated by the local controllers. $y^{\star}$ denotes the reference signal for the central controller. The reference signal for the local controller is set to zero because disturbance rejection is desired at high frequency. LPF, and HPF represents the low pass and high pass filters.
% \begin{figure} [htbp]
%     \centering
%     \includegraphics[scale=0.5]{img.pdf}
%     \caption{Power park system model.}
%     \label{fig:fr_pt}
%     %\vspace{-0.15in}
% \end{figure}

Our test system has three inverters, where each inverter contains an inner current control loop embedded within an outer voltage control loop. A 5G network is used to communicate the control signals from the central controller to the individual local controllers at each inverter. The inverters and loads are connected to the generator bus. Further details about this case can be found in \cite{1705638}.
 
% \begin{figure} [htbp]
%     \centering
%     \vspace{0.10in}

%     \includegraphics[width=\columnwidth]{pp2.pdf}
%     \caption{Frequency partitioning in inverter control system \cite{1705638}.}
%     \label{fig:cs_pp}
%     \vspace{-0.10in}
% \end{figure}

\subsection{Performance Evaluation}

\subsubsection{Set point change in voltage at the generator bus}
\begin{figure} [htbp]
    \centering
    \includegraphics[width=\columnwidth]{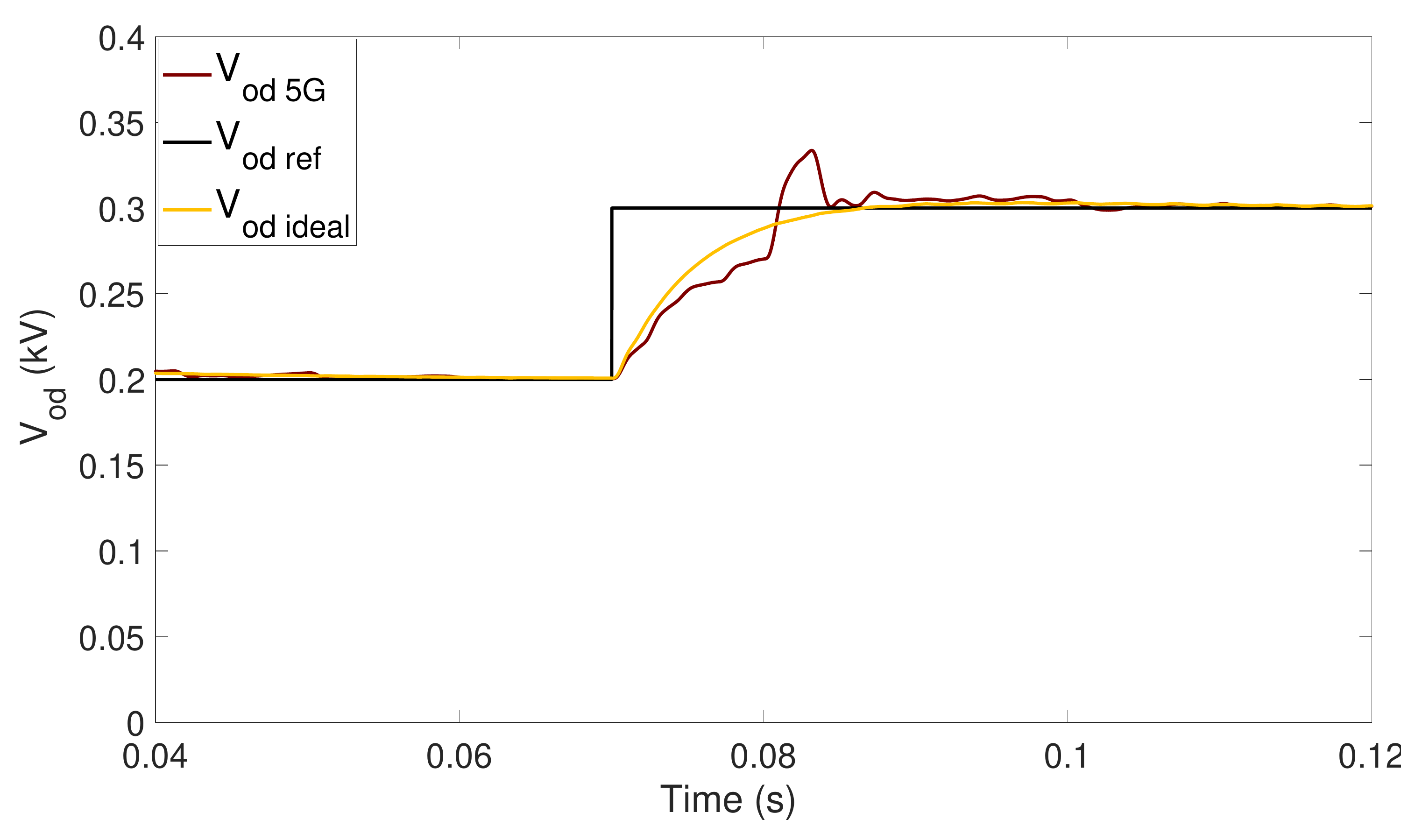}
    \caption{Comparison of the system response ($d$-component of the voltage at the generator bus) to a step change in set-point with and without the 5G network.}
    \label{fig:pp_res}
\end{figure}

This case study evaluates voltage regulation at the common generator bus. The analysis is carried out in the $dq$-frame of reference. $V_{oq}$ is set to zero and a step change is applied in $V_{od}$ from 0.2 to 0.3 at $t=0.07~\text{s}$. The system response is shown in  Fig.~\ref{fig:pp_res}. $V_{\text{od-ideal}}$ denotes the system simulation response under ideal conditions (5G channel is not used). $V_{\text{od-5G}}$ denotes the system response with 5G communication channel (the low frequency control signal is communicated from the central controller to the individual local controllers). Table~\RN{2}shows the overshoot and settling time of the responses. The settling times for both the cases are around $0.01~\text{s}$. There is, however, a change in overshoot from 1\% to 10.3\% in the presence of the 5G network. 
    
\begin{table}[htbp]
    \centering
    \caption{Overshoot and Settling time for set point change in voltage at the generator bus (Power park case) }
    \label{table:pp_perform}
    \scriptsize
    %  \begin{tabular}{|p{3cm}|p{3.6cm}|}
    \begin{tabular}{lll}
    \toprule
    \textbf{Case}  & \textbf{Overshoot (\%)} & \textbf{Settling Time (ms)} \\ \midrule
    Ideal  & 1 & 11.1 \\ 
    5G & 10.3 & 14.19 \\ \bottomrule
    \end{tabular}
\end{table}

\subsubsection{Load switching}
\begin{figure} [tb]
    {\includegraphics[width=\columnwidth]{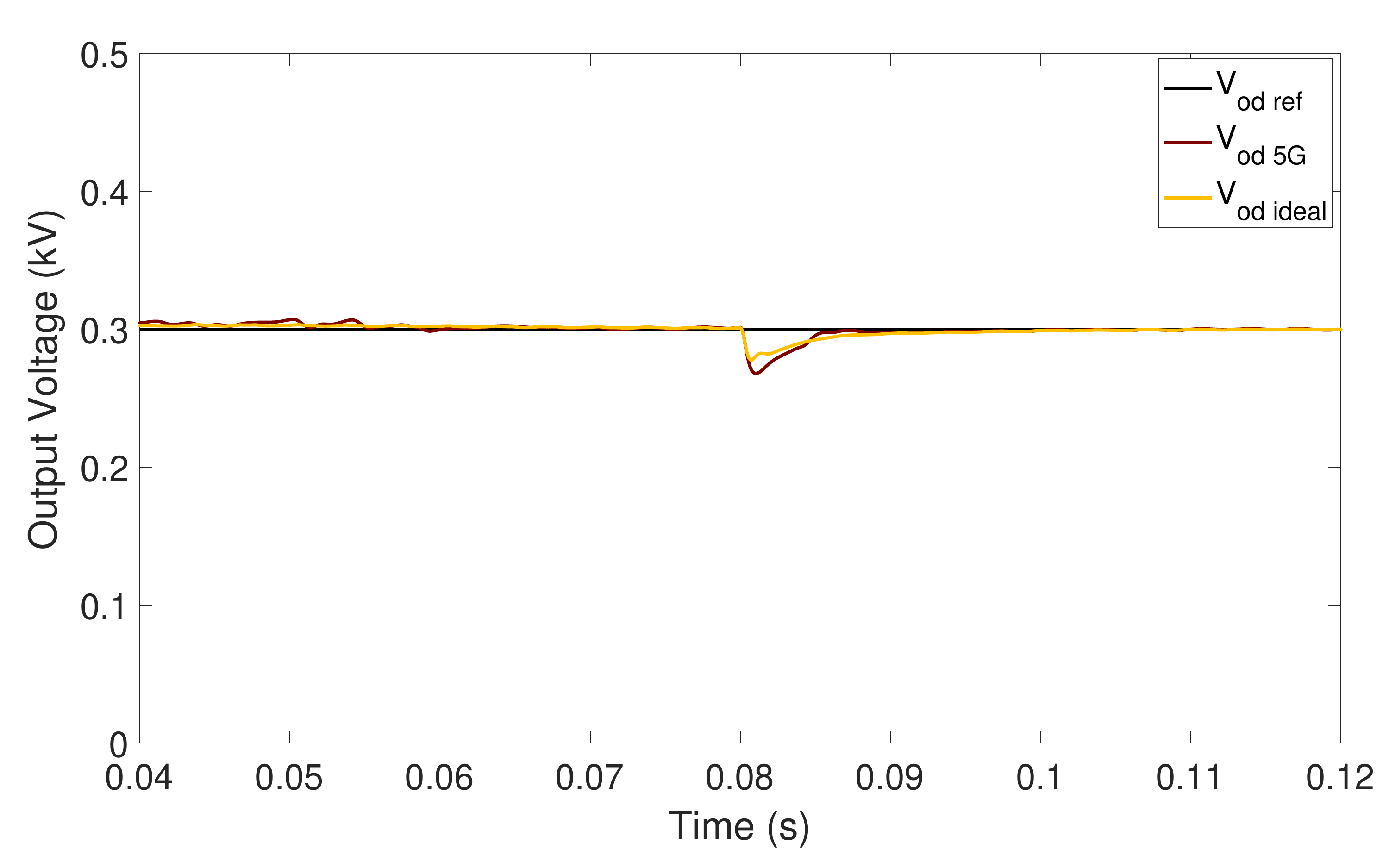}\label{l_sw_vd.pdf}}   
    \caption{Load switching of a three phase balanced RL load. Output voltage in $dq$ frame ($d$-axis component).}
    \label{fig:l_sw}
\end{figure}

The system with and without 5G communication network is subjected to a load switching condition. A three-phase balanced RL load is switched on and is connected to the system at $t= 0.08~\text{s}$. The transient response is shown in Fig.~\ref{fig:l_sw}. This includes the $d$-axis component of the output voltage in $dq$-frame of reference. The results indicate similar transient performance for both cases. The transient lasts for around $5~\text{ms}$ in both the cases and the system does not lose its stability. The drop in voltage is around $0.01~\text{V}$ more with the 5G channel. 

\subsubsection{Single-phase to ground fault at the generator bus}
\begin{figure}[htbp]
\centering
{\includegraphics[width=\columnwidth]{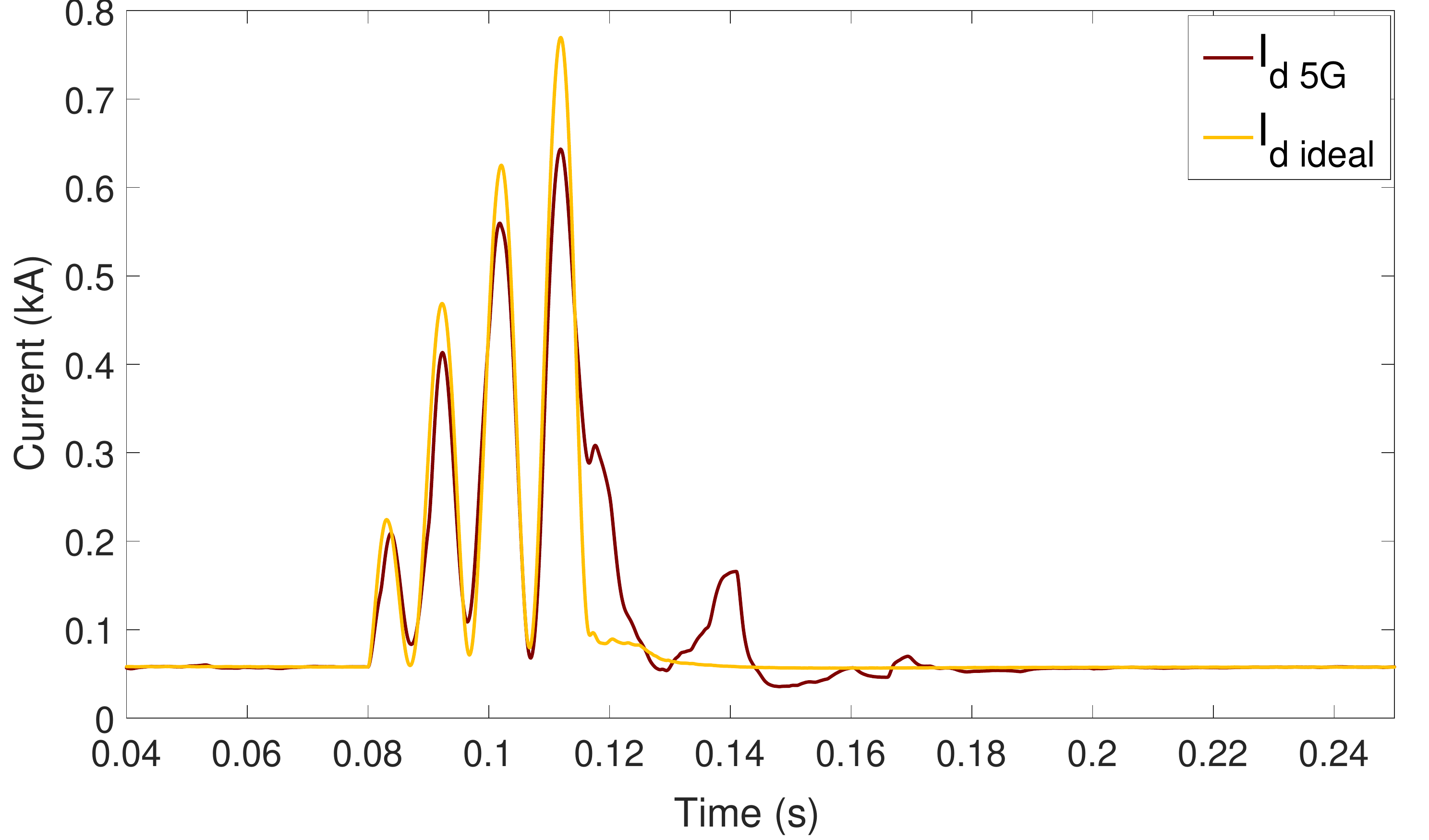}\label{id_1f}}
(a)
{\includegraphics[width=\columnwidth]{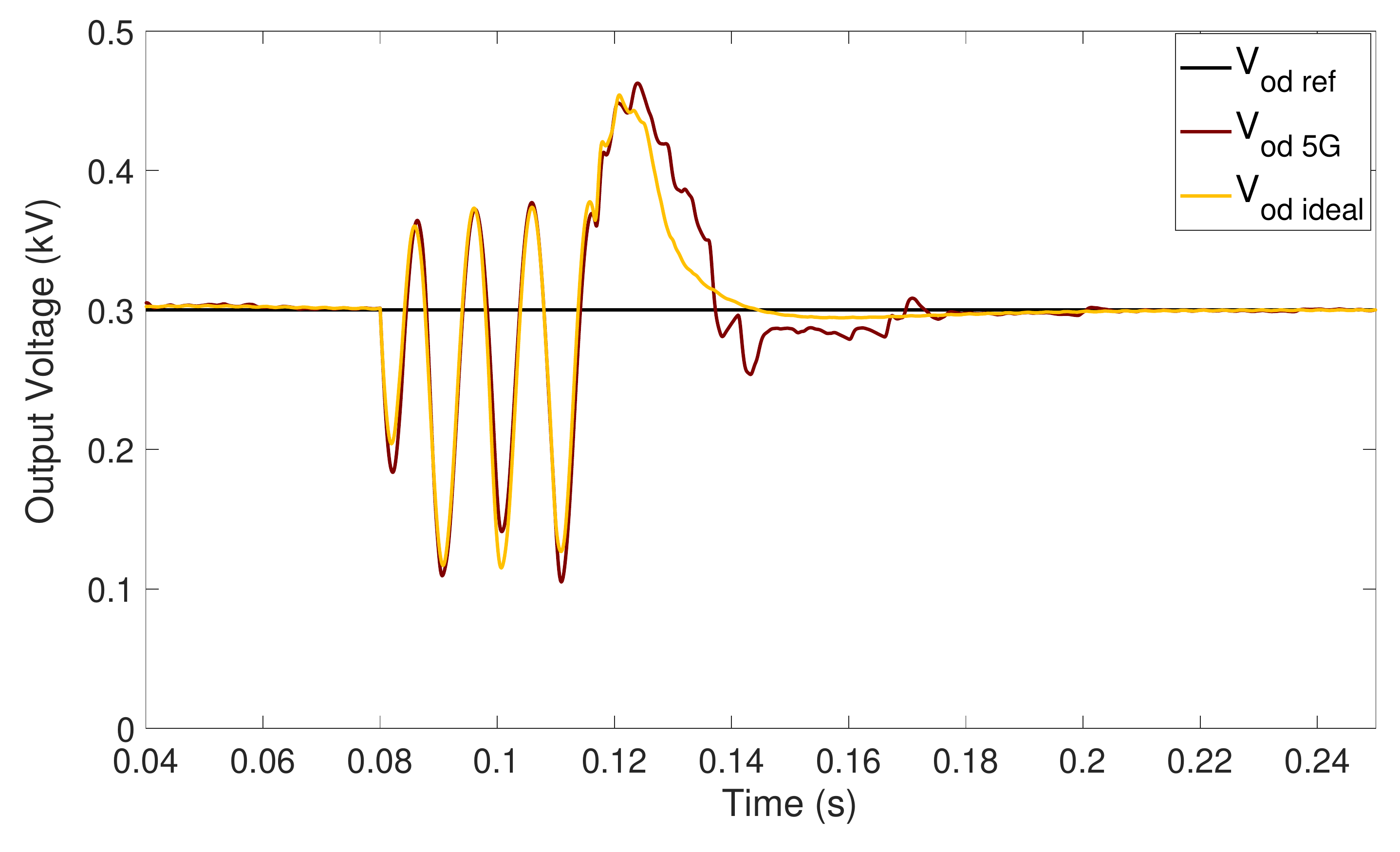}\label{vd_1f}}
(b)
\caption{Single-phase fault to ground (a) $d$-component of the output current of one of the inverters in $dq$-frame. (b) $d$-component of the output voltage at the generator bus in $dq$-frame.}
\label{1f}
\end{figure}

A single-phase A-G fault is simulated at $t = 0.08~\text{s}$ for $0.03~\text{s}$ at the generator bus. Fig.~\ref{1f} shows the results. The system with 5G network does not lose its stability. It is able to continue tracking the reference value of the direct axis voltage at the generator bus after the fault is cleared. The transient lasts for about $0.065~\text{s}$ after the fault is cleared with 5G channel, while it lasts for around $0.028~\text{s}$ with ideal communication.
 
\subsubsection{Three-phase to ground fault at the generator bus}
\begin{figure}[htbp]
\centering
\includegraphics[width=\columnwidth]{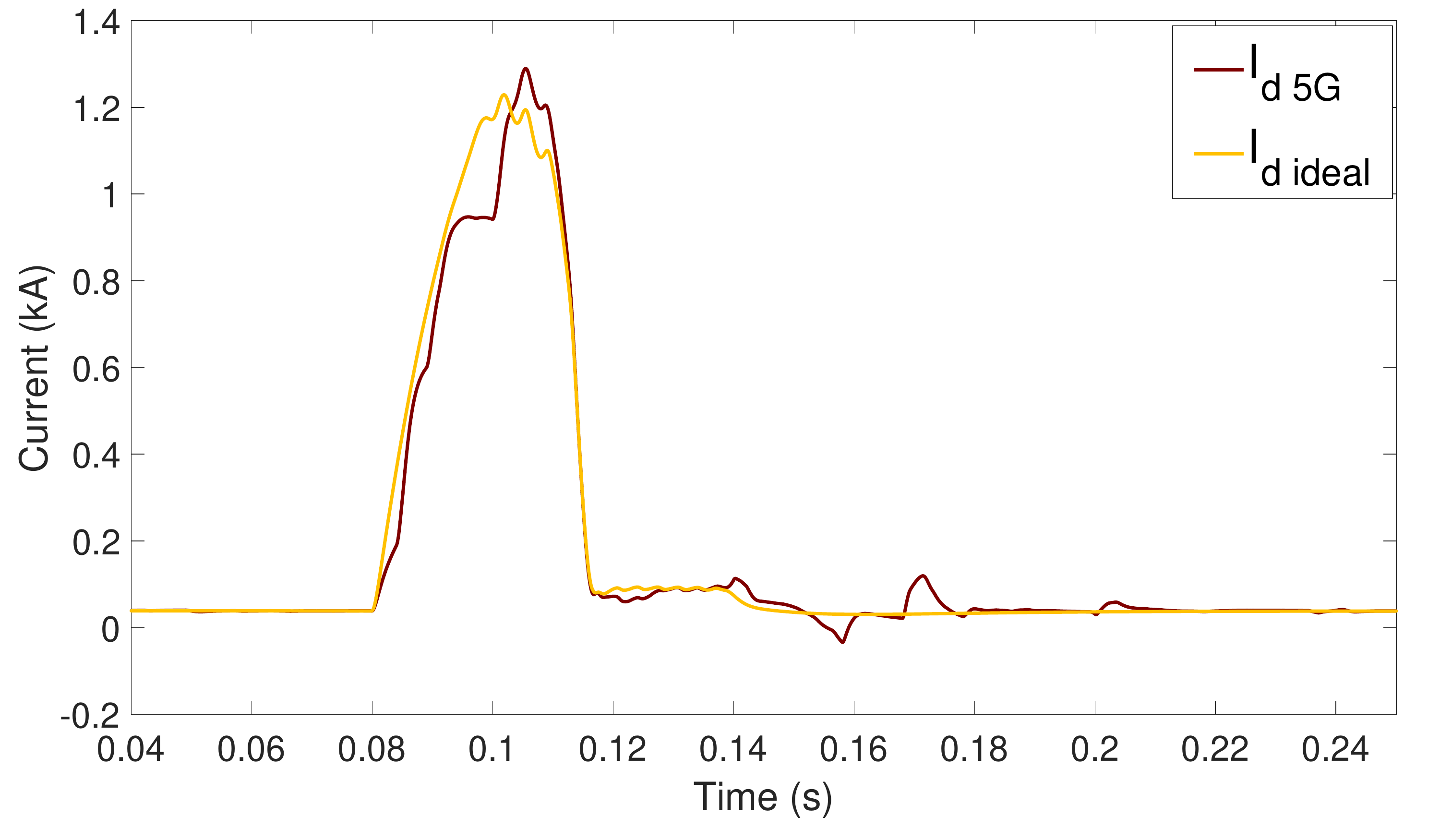}\label{id_3f}
(a)
\includegraphics[width=\columnwidth]{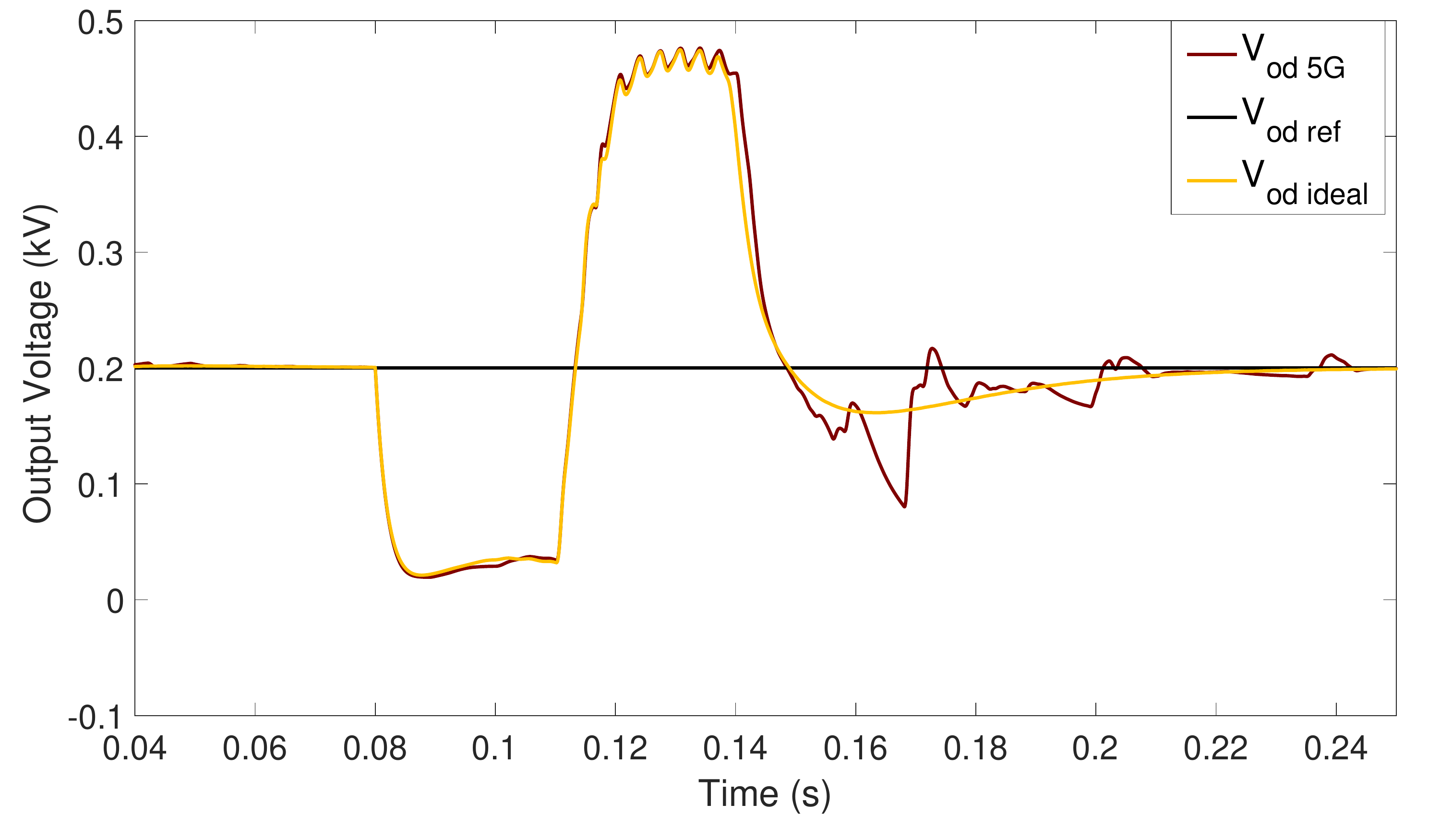}\label{vd_3f}
(b)
\caption{Three-phase fault to ground (a) $d$-component of the output current of one of the inverters in $dq$-frame. (b) $d$-component of the output voltage at the generator bus in $dq$-frame.}
\label{3f}
\end{figure}

A three-phase ABC-G fault is simulated at $t = 0.08~\text{s}$ for $0.03~\text{s}$ at the generator bus. Fig.~\ref{3f} shows the results. The system with 5G network does not lose its stability. It is able to continue tracking the reference value of the direct axis voltage at the generator bus after the three-phase to ground fault is cleared. The transient lasts for 
%about 
$0.13~\text{s}$ after the fault is cleared with 5G channel, while it lasts for %around
$0.1~\text{s}$ with ideal communication.
 
%% file: 5-cspm.tex
\section{Use Case II: Coordinated Set Point Modulation of DERs}
\label{sec:cspm}
\subsection{System Description}
Set point modulation involves modulating the system set-point $x_{sp}$ depending on the system response $x(t)$~\cite{6179568}. Mathematically, this is given as
\begin{equation}
    x_{sp}^{\prime}= x_{sp}+me(t),
\end{equation}
where $x_{sp}^{\prime}$ is the modulated set point, $m$ is a design parameter, and $e(t)$ is the tracking error that measures the deviation of of the system response from the set point.
\begin{equation}
    e(t)= x_{sp} - x(t).
\end{equation}
For better dynamic response, the tracking error $e(t)$ is replaced by the the predictive dynamic behaviour of the tracking error. That is,
\begin{equation}
    x_{sp}^{\prime}= x_{sp}+m{\hat{e}}_{pred}(t).
\end{equation}
There are different methods that can be used to obtain ${\hat{e}}_{pred}(t)$, but this case uses a linear expression.

The above equations are for a single DER. Another level of control can be added with the help of communication between different DERs known as coordinated set point modulation~\cite{cspmref}. Consider $N$ devices connected at the PCC that are capable of exchanging information over a communication channel. The devices connected through a communication link exchange information about their respective predictive tracking errors.
% Each device exchanges information about its predictive tracking error with the devices to which it is connected over the communication space.
The existing equation for a single DER can be modified to represent the system response for the $i$th DER as
\begin{equation}
      x_{{i}_{sp}}^{\prime}= x_{i_{sp}}+m_{i}{\hat{e}}_{i_{pred}}(t)
\end{equation}
The secondary level of control $u(t)$ can then be defined as:
\begin{equation}
     u(t)= m_{i}\sum_{j=1,j\neq i}^Na_{ij}{\hat{e}}_{j_{pred}}(t)
\end{equation}
where $a_{ij}$ denotes a communication link between DER $i$ and DER $j$. Hence the control equation can be represented by
\begin{equation}
      x_{{i}_{sp}}^{\prime}=x_{i_{sp}}+m_{i}{\hat{e}}_{i_{pred}}(t)+ m_{i}\sum_{j=1,j\neq i}^Na_{ij}{\hat{e}}_{j_{pred}}(t)
      \label{eqn:cspm}
\end{equation}

Further details about this concept can be found in \cite{cspmref}. 
\subsection{Performance Evaluation}
\begin{figure} [htb]
    \centering
    \includegraphics[scale=0.4]{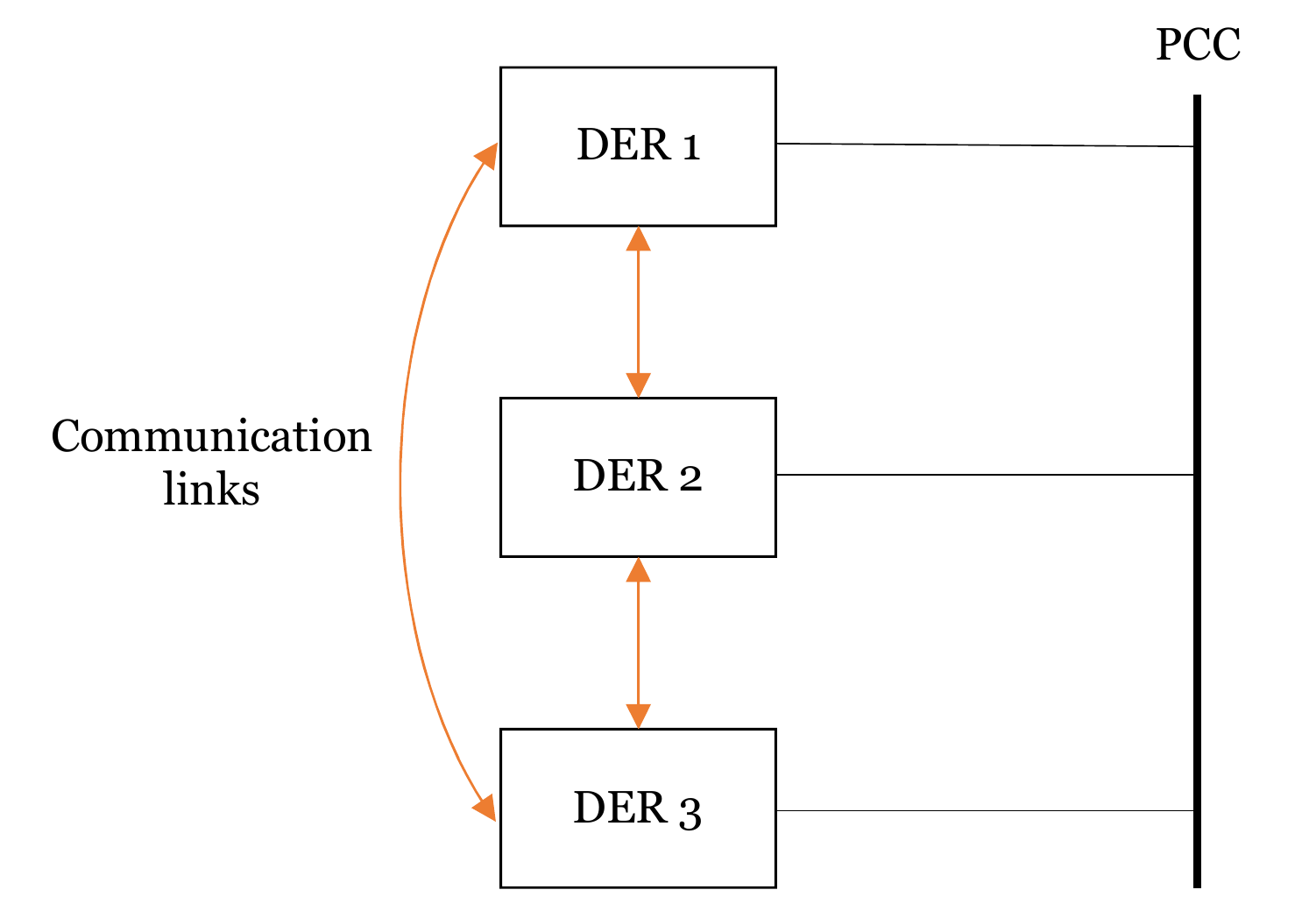}
    \caption{System model of coordinated set point modulation of DERs.}
    \label{fig:cspm_model}
    % \vspace{-0.15in}
\end{figure}

Fig.~\ref{fig:cspm_model} shows the three devices (DERs) that interact with each other using %the concept of
coordinated set-point modulation. 

\subsubsection{Staggered set point change}
\begin{figure} [htbp]
%\vspace{0.15 in}
%\centering
\includegraphics[width=\columnwidth]{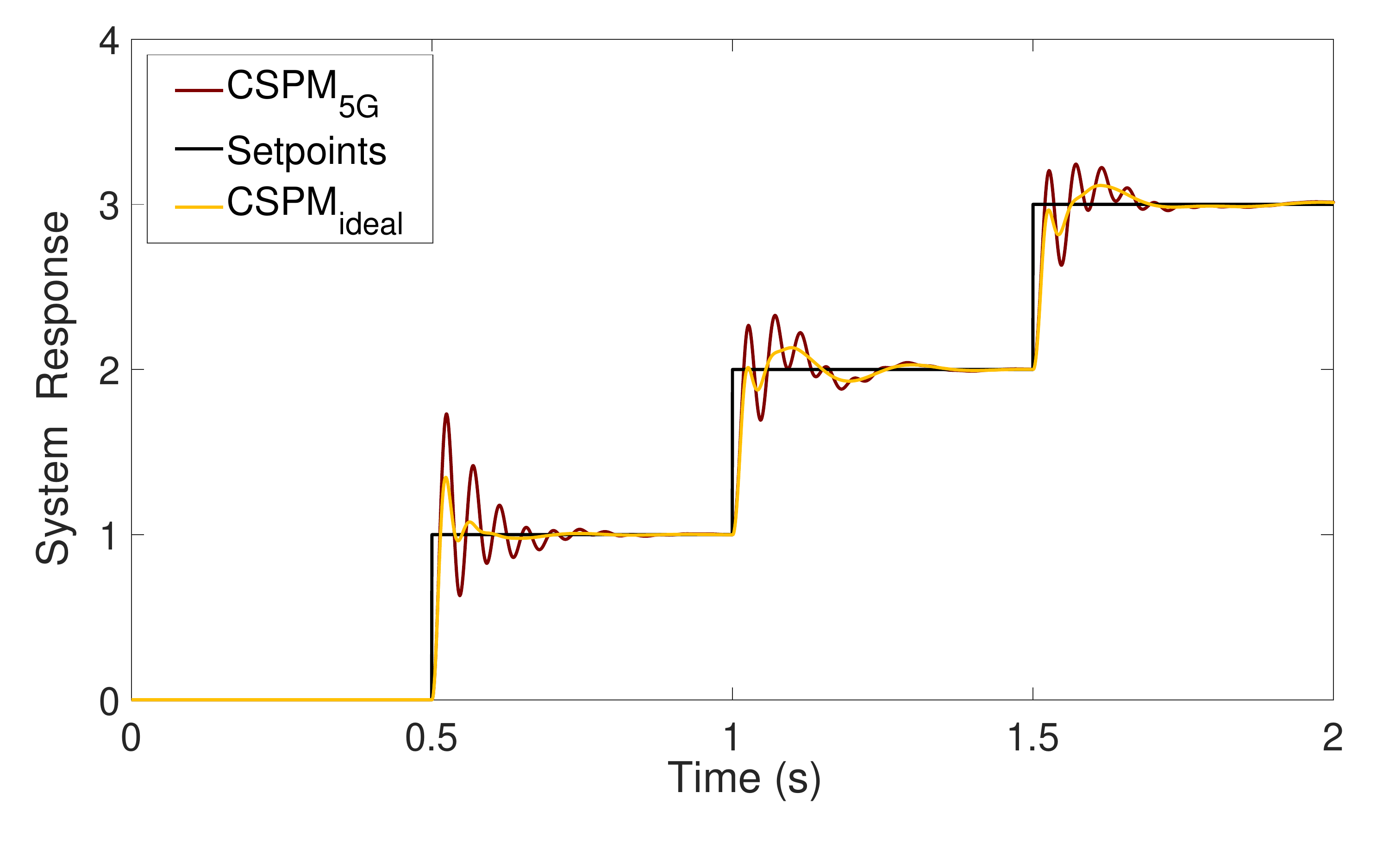}
%\vspace{0in}
\caption{Comparison of system response for coordinated set point modulation (staggered set point change) with and without 5G network.}
\label{fig:csp_res_stag}
% \vspace{-0.05 in}
\end{figure}

A staggered set point change is applied to the individual devices (step change in reference point from 0 to 1 at $t = 0.5~\text{s}$ for DER 1, at $t = 1~\text{s}$ for DER 2 and at $t = 1.5~\text{s}$ for DER 3). These set points can represent any electrical quantity that needs control (voltage, real power or reactive power). The combined system response at the PCC, under ideal simulation conditions (no 5G communication involved) and in the presence of a 5G network, is shown in Fig.~\ref{fig:csp_res_stag}. Table~\RN{3}shows the overshoot and the settling time of the system response under the two conditions. The settling time increases by $172~\text{ms}$ in the presence of a 5G network and the overshoot is twice the ideal simulation condition. However, the DERs are able to track the set point changes successfully with 5G communication.    

\subsubsection{Simultaneous set point change}
\begin{figure} [htbp]
%\vspace{0.15 in}
%\centering
\includegraphics[width=\columnwidth]{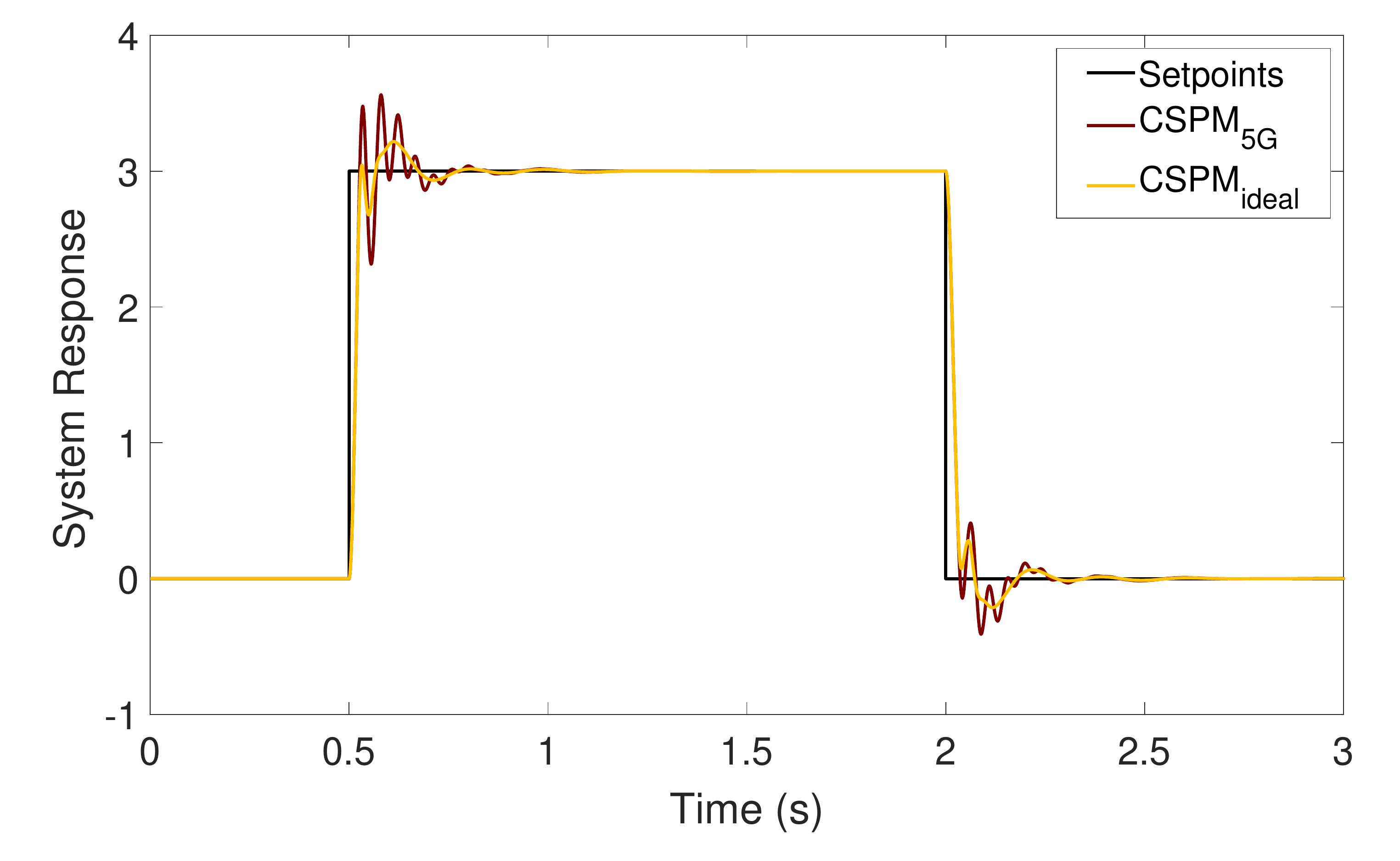}
%\vspace{0in}
\caption{Comparison of system response for coordinated set point modulation (simultaneous set point change) with and without the 5G network.}
\label{fig:csp_res_sim}
% \vspace{-0.05 in}
\end{figure}

A simultaneous set point change is applied to all the devices (step change in reference point from 0 to 1 at $t=0.5~\text{s}$). The set point is brought back to 0 for all the devices at $t=2~\text{s}$. The combined system response at the PCC, under ideal simulation conditions and in the presence of a 5G network, is shown in Fig.~\ref{fig:csp_res_sim}. Table~\RN{3}shows the overshoot and the settling time of the system response under the two conditions. The settling time increases by $84~\text{ms}$ in the presence of a 5G network and the overshoot is higher by 11.4\%. However, the set points are tracked successfully without loss in stability.

\subsubsection{Communication failure in one of the devices}
\begin{figure} [htbp]
%\vspace{0.15 in}
%\centering
\includegraphics[width=\columnwidth]{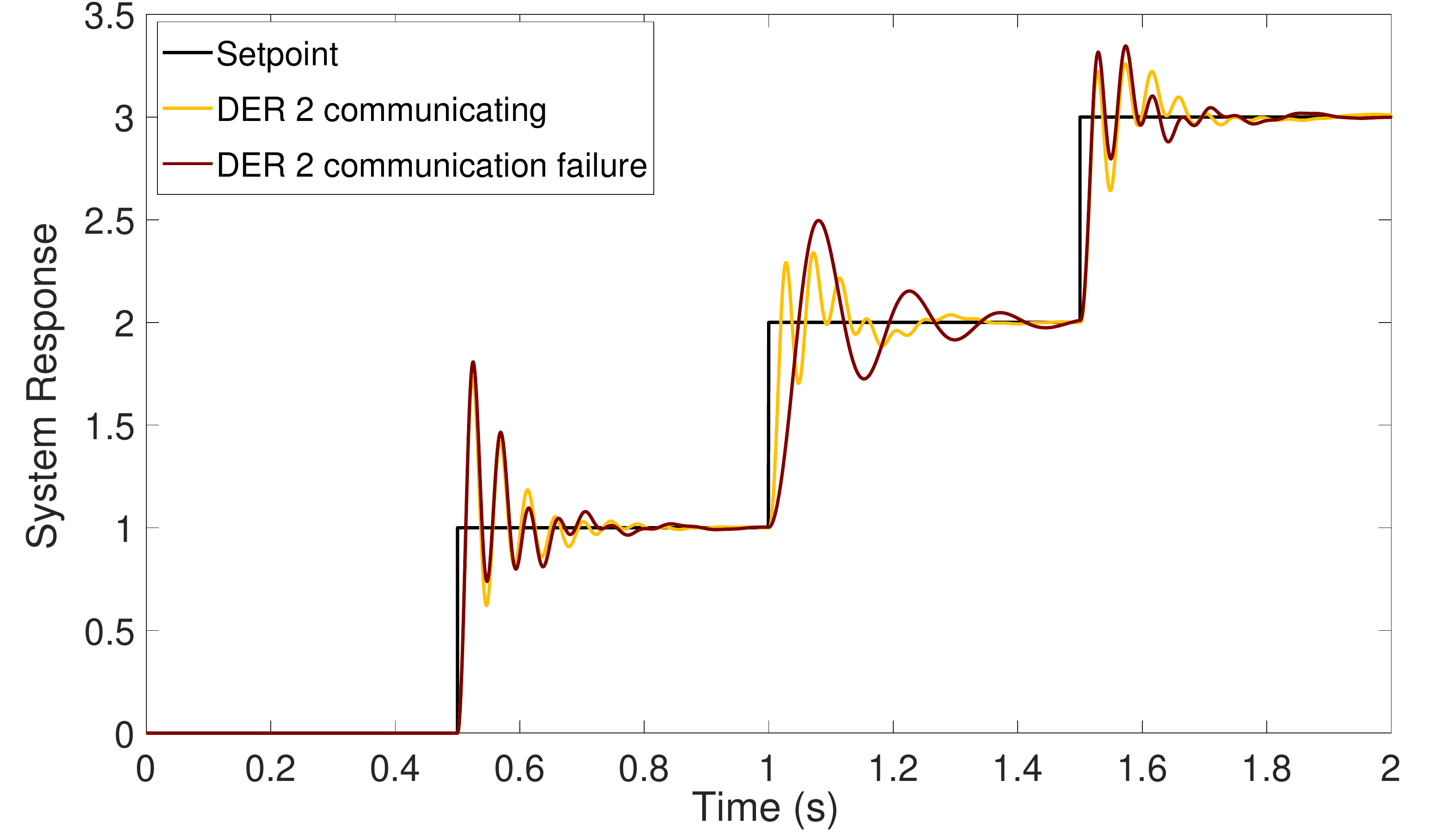}
%\vspace{0in}
\caption{Comparison of system response for coordinated set point modulation (staggered set point change) under ideal communication conditions and when communication for device 2 fails.}
\label{fig:cspm_cf_2}
% \vspace{-0.05 in}
\end{figure}

The system performance is evaluated when one of the devices is incapable of communicating its state to the other two devices. The response is shown in Fig.~\ref{fig:cspm_cf_2}. The response is also compared with the scenario in which all the devices can communicate without any hindrance. Although there is an increase in overshoot and settling time, the system is able to track the set points owing to the control algorithm used. This can be seen from the first two terms in \eqref{eqn:cspm}. These terms do not depend on the states of the other devices, thus contributing towards set point tracking in the absence of information from one device.    

\begin{table}[H] % htbp
\centering
\vspace{0 in}
\caption{System Performance for set point changes (Coordinated set point modulation case)}
 \label{table:cspm_r}
%  \begin{tabular}{|p{3cm}|p{3.6cm}|}
    \begin{tabular}{p{8 em}lll l}
    \toprule
    \textbf{Set point change} & \textbf{Case}  & \textbf{Overshoot (\%)} & \textbf{Settling Time (ms)} \\ \midrule
     Staggered & Ideal  & 34 & 78 \\ 
    & 5G & 73.1 & 250 \\ \midrule
    Simultaneous & Ideal  & 7.2 & 153 \\ 
    & 5G & 18.6 & 237 \\ \bottomrule
\end{tabular}
\vspace{0 in}
\end{table}

%% file: 6-conclusion.tex
\section{Conclusion}
\label{sec:conclusion}

In this work, the effects of delays induced by 5G scheduling on the performance of smart grids have been investigated. Results from the two cases considered show that the grid is able to track the set point changes and handle transients in the presence of a 5G communication network. There is a slight degradation in the system performance as compared to ideal simulation values without the presence of a 5G network, but the system still maintains stability. In the future, we will explore different RB scheduling policies like best channel quality indicator, proportional fairness and analyzing their effect on system performance. 

%% best cqi - https://eudl.eu/pdf/10.4108/icst.5gu.2014.257987

% The current simulation does not take into account the CN, which is one of the most critical part of a 5G network. While the delay and reliability between the RAN and CN can be hard-coded, a realistic simulation needs to incorporate it to get the full picture. Additionally in the current simulation, each device generates traffic in the same manner which may not hold for specific cases, as devices generate traffic based on the application which may not be the same across multiple devices. Finally, a real-world smart grid is likely to have multiple devices and a large scale simulator should be developed to account for it.

% Based on the initial results, the following aspects will be investigated in the future 

% \begin{itemize}
%     \item Incorporating a more realistic 5G simulator with both the radio access network and the network core. % operating on top of the existing simulation.
%     \item Investigating different RB scheduling policies like best channel quality, proportional fairness and analysing their effect on system performance.
% \end{itemize}